\documentclass[manuscript]{aastex}
\usepackage{amsmath}
\usepackage{graphicx}

\shorttitle{Carbon Isotope and Isotopomer Fractionation}
\shortauthors{Furuya et al.}

\begin{document}


\title{Carbon Isotope and Isotopomer Fractionation \\in Cold Dense Cloud Cores}


\author{Kenji Furuya\altaffilmark{1}, Yuri Aikawa\altaffilmark{1}, Nami Sakai\altaffilmark{2} and Satoshi Yamamoto\altaffilmark{2}}
\affil{\altaffilmark{1}Department of Earth and Planetary Sciences, Kobe University, Kobe 657-8501, Japan}
\affil{\altaffilmark{2}Department of Physics and Research Center for the Early Universe, The University of Tokyo, Bunkyo-ku, Tokyo 113-003, Japan}
\email{furuya@stu.kobe-u.ac.jp}


%\altaffiltext{1}{Visiting Astronomer, Cerro Tololo Inter-American Observatory.
%CTIO is operated by AURA, Inc.\ under contract to the National Science
%\altaffiltext{2}{Society of Fellows, Harvard University.}
%\altaffiltext{3}{present address: Center for Astrophysics,
%    60 Garden Street, Cambridge, MA 02138}
%\altaffiltext{4}{Visiting Programmer, Space Telescope Science Institute}
%\altaffiltext{5}{Patron, Alonso's Bar and Grill}


\begin{abstract}

We construct the gas-grain chemical network model which includes carbon isotopes ($^{12}$C and $^{13}$C) with an emphasis on isotopomer-exchange reactions. Temporal variations of molecular abundances, the carbon isotope ratios ($^{12}$CX/$^{13}$CX) and the isotopomer ratios ($^{12}$C$^{13}$CX/$^{13}$C$^{12}$CX) of CCH and CCS in cold dense cloud cores are investigated by numerical calculations. We confirm that the isotope ratios of molecules, both in the gas phase and grain surfaces, are significantly different depending on whether the molecule is formed from the carbon atom (ion) or the CO molecule. Molecules formed from carbon atoms have the CX/$^{13}$CX ratios greater than the elemental abundance ratio of [$^{12}$C/$^{13}$C]. On the other hand, molecules formed from CO molecules have the CX/$^{13}$CX ratios smaller than the [$^{12}$C/$^{13}$C] ratio. We reproduce the observed C$^{13}$CH/$^{13}$CCH ratio in TMC-1, if the isotopomer exchange reaction, $^{13}$CCH + H $\rightleftharpoons$ C$^{13}$CH + H + 8.1 K, proceeds with the forward rate coefficient $k_f$ $>$ 10$^{-11}$ cm$^3$ s$^{-1}$. However, the C$^{13}$CS/$^{13}$CCS ratio is lower than that observed in TMC-1. We then assume the isotopomer exchange reaction catalyzed by the H atom, $^{13}$CCS + H $\rightleftharpoons$ C$^{13}$CS + H + 17.4 K. In the model with this reaction, we reproduce the observed C$^{13}$CS/$^{13}$CCS, CCS/C$^{13}$CS and CCS/$^{13}$CCS ratio simultaneously.
%and C$^{13}$CH + S $\rightarrow$ $^{13}$CCS + H lowers the C$^{13}$CS/$^{13}$CCS ratio.

%We also find that observed $^{12}$C$^{13}$CH/$^{13}$C$^{12}$CX ratio and $^{12}$C$^{13}$CS/$^{13}$C$^{12}$CS ratio are possibly made through the exchange of the $^{13}$C position after formation of molecules. 
%it is found that observed ratios cannot be explained by    
\end{abstract}





\section{INTRODUCTION}

The elemental abundance ratio of [$^{12}$C/$^{13}$C] in the local ISM is 60 (e.g. Langer \& Penzias (1993); Lucas \& Liszt (1998)). The molecular abundance ratio of $^{12}$CX/$^{13}$CX is often assumed to be the same as this elemental ratio. It has been reported, however, that the carbon-chain molecules are diluted in $^{13}$C and the ratio varies among molecules; the ratio of HC$_3$N/H$^{13}$CCCN is 79 (Takano et al. 1998), CCS/$^{13}$CCS is 230 $\pm$ 130 (Sakai et al. 2007) and CCH/$^{13}$CCH is larger than 250 (Sakai et al. 2010). These ratios should reflect the production pathway of each molecule. In addition, recent observations in TMC-1 indicated that the abundances of the $^{13}$C isotopomers depend on which carbon atom in a molecule is substituted by $^{13}$C. Takano et al. (1998) observed the three $^{13}$C isotopomers of HC$_3$N, and found that HCC$^{13}$CN is more abundant than HC$^{13}$CCN and H$^{13}$CCCN, which indicates three carbon atoms are not equivalent in HC$_3$N. Takano et al. (1998) suggested that the neutral-neutral reaction of C$_2$H$_2$ + CN is the main formation path to HC$_3$N. Sakai et al. (2007; 2010) reported the abundance ratios of C$^{13}$CS/$^{13}$CCS = 4.2 $\pm$ 2.3 and C$^{13}$CH/$^{13}$CCH = 1.6 $\pm$ 0.4. Again, these results indicate two carbon atoms are not equivalent in CCS and CCH. Sakai et al. (2007; 2010) suggested that the neutral-neutral reactions of CH + CS and CH$_2$ + C would significantly contribute to the CCS and CCH formation, respectively. These results indicate the neutral-neutral reactions might play an important role in production of such carbon-chain molecules. In this way, observations of $^{13}$C species would be useful to investigate chemistry of organic species.
   
Langer et al. (1984) and Langer \& Graedel (1989) constructed theoretical models of the $^{13}$C chemistry in the gas phase. They showed that carbon bearing species are divided into three groups: CO, HCO$^+$ and the "carbon isotope pool". CO is $^{13}$C-rich and "carbon isotope pool" is $^{13}$C-poor. HCO$^+$, on the other hand, can be either $^{13}$C-rich or $^{13}$C-poor. Recently Woods \& Willacy (2009) revisited carbon isotope fractionation including the adsorption of gaseous molecules onto grain surfaces and grain surface reactions, but they do not distinguish $^{13}$C isotopomers. In addition, temporal evolutions in molecular clouds are not investigated in detail, because they focused on $^{13}$C species in protoplanetary disks.

In this paper, we present a detailed chemical model in cold dense cloud cores to investigate carbon isotope fractionation and isotopomer fractionation of CCH and CCS. In Section 2, we describe our fiducial model. In Section 3, we show the evolutions of molecular abundances, isotope ratios and isotopomer ratios in the fiducial model. In Section 4, we compare our results with the observations in TMC-1. We also discuss the dependence of our results on the assumptions of rate coefficients and initial condition of carbon. Since some terms are confusing, we define them here; "carbon isotope ratio" is the abundance ratio of the $^{12}$CX/$^{13}$CX, and "carbon isotopomer ratio" is the abundance ratio of the $^{12}$C$^{13}$CX/$^{13}$C$^{12}$CX. If a carbon isotope ratio deviates from the elemental [$^{12}$C/$^{13}$C] ratio, we call it "carbon isotope fractionation". If carbon isotopomer ratio deviates from unity, we call it "carbon isotopomer fractionation".   

\section{FIDUCIAL MODEL}

%% In a manner similar to \objectname authors can provide links to dataset
%% hosted at participating data centers via the \dataset{} command.  The
%% second curly bracket argument is printed in the text while the first
%% parentheses argument serves as the valid data set identifier.  Large
%% lists of data set are best provided in a table (see Table 3 for an example).
%% Valid data set identifiers should be obtained from the data center that
%% is currently hosting the data.
%%
%% Note that AASTeX interprets everything between the curly braces in the 
%% macro as regular text, so any special characters, e.g. "#" or "_," must be 
%% preceded by a backslash. Otherwise, you will get a LaTeX error when you 
%% compile your manuscript.  Special characters do not 
%% need to be escaped in the optional, square-bracket argument.   
\subsection{Carbon isotope fractionation}

There are two processes which fractionate carbon isotopes in molecular clouds: selective photodissociation and isotope-exchange reactions. 

\subsubsection{Selective photodissociation}

Carbon monoxide is photodissociated mostly by line absorption. Because of the high abundance of CO, self-shielding is efficient especially for $^{12}$CO. Therefore the $^{12}$CO/$^{13}$CO ratio is enhanced in the very thin surface region ($A_V$ $<$ 1 mag) (e.g. van Dishoeck et al. 1988). In this study we consider central regions of molecular cloud cores with $A_V$ $\sim$ 10 mag, and hence we do not consider this process in the following sections.

%The absorption line of $^{12}$CO is saturated at cloud surface, while $^{13}$CO is photo dissociated in deeper regions. 
\subsubsection{Fractionation through isotope-exchange reactions}

There are small zero-point vibrational energy differences among the isotopically distinct species. In a cloud of low temperature ($T \sim$ 10 K) the isotope fractionation occurs through isotope-exchange reactions. Watson et al. (1976) pointed out that the carbon isotope fractionation in interstellar molecules mainly occurs as a result of the reaction:
%There is small zero-point energy difference between the $^{12}$C specie and the $^{13}$C specie.If the cloud temperature is near absolute zero, isotope fractionation may occur through isotope exchange reactions. Most important exchange reaction is     
\begin{equation}
   ^{13}{\rm C}^{+} + {\rm ^{12}CO} \rightleftharpoons {\rm ^{13}CO} + {\rm ^{12}C}^{+} + \Delta E,
\end{equation}
\noindent
where $\Delta E$, the zero-point vibrational energy difference between the reactants and products, is 35 K. The rate coefficient of this reaction was measured by Watson et al. (1976) at 300 K and by Smith \& Adams (1980) at 80 - 510 K. Langer et al. (1984) extrapolated these experimental data below 80 K assuming the following relation between the forward reaction rate coefficient $k_f$ and the backward reaction rate coefficient $k_r$:  
\[
k_r = k_f \hspace{3pt}{\rm exp}(-\Delta E/k_{\rm B}T),
\]
where $k_{\rm B}$ is Boltzmann coefficient. At 10 K, the forward rate coefficient is 1.34 $\times$ 10$^{-9}$ cm$^3$ s$^{-1}$ and the backward rate coefficient is 4.0 $\times$ 10$^{-11}$ cm$^3$ s$^{-1}$. Smith \& Adams (1980) also measured the rate coefficient of the following proton transfer reaction at 80 - 510 K:  
%as first pointed out by Watson et al.(1976).Langer et al.(1984) calculated forward reaction rate, 1.34$\times$10$^{-9}$cm$^{-3}$ and backward reaction rate 4.0$\times$10$^{-9}$cm$^{-3}$. Other exchange reaction is
\begin{equation}  
   {\rm H}{\rm ^{12}CO}^{+} + {\rm ^{13}CO} \rightleftharpoons {\rm ^{12}CO} + {\rm H}{\rm ^{13}CO}^{+} + \Delta E,
\end{equation}
\noindent
where the vibrational zero-point energy difference is $\Delta E$ = 9 K. Langer et al. (1984) estimated the reaction rate coefficient at 10 K: $k_f$ = 6.5 $\times$ 10$^{-10}$ cm$^3$ s$^{-1}$ and $k_r$ = 2.7 $\times$ 10$^{-10}$ cm$^3$ s$^{-1}$.
 
\subsection{Carbon isotopomer fractionation}

Sakai et al. (2007; 2010) discussed the cause of carbon isotopomer fractionation of CCS and CCH based on their observational results in TMC-1. They pointed out two possibilities: (i) the formation pathways of the species and (ii) the exchange of the $^{13}$C position after formation of molecules by isotopomer-exchange reactions. Specifically they proposed the reactions in Table 1. 

\subsubsection{Fractionation through formation paths}
If the observed fractionation is due to the formation paths, the following neutral-neutral reaction (e.g. Turner et al. 2000) could significantly contribute to the CCH formation:
%If the former is the main process of the isotopomer fractionation, the major production pathway of the molecule must not include symmetric molecules. Based on this idea, they determined the major production reactions of CCH and CCS under the assumption that the effect of (ii) is negligible. 
\[
  {\rm C} + {\rm CH}_2 \rightarrow {\rm CCH} + {\rm H}.
\]  
\noindent
Two carbon atoms are not equivalent in this neutral-neutral reaction:
\begin{equation}
  {\rm ^{12}C} + {\rm ^{13}CH}_2 \rightarrow {\rm ^{12}C}{\rm ^{13}CH} + {\rm H}, 
\end{equation}
\begin{equation}
  {\rm ^{13}C} + {\rm ^{12}CH}_2 \rightarrow {\rm ^{13}C}{\rm ^{12}CH} + {\rm H}.
\end{equation}
\noindent
If the isotope ratios of the carbon atom and CH$_2$ are different, the isotopomer ratio of CCH deviates from unity. 

Similarly, the observed isotopomer fractionation of CCS could be due to the neutral-neutral reaction:
\[
  {\rm CH} + {\rm CS} \rightarrow {\rm CCS} + {\rm H},
\]  
\noindent
in which two carbon atoms are not equivalent:
\begin{equation}
  {\rm ^{12}CH} + {\rm ^{13}CS} \rightarrow {\rm ^{12}C}{\rm ^{13}CS} + {\rm H},
\end{equation}
\begin{equation}    
  {\rm ^{13}CH} + {\rm ^{12}CS} \rightarrow {\rm ^{13}C}{\rm ^{12}CS} + {\rm H}.
\end{equation}

In addition, the isotopomer fractionation of CCH might propagate to CCS through the following reactions (Yamada et al. 2002):
\begin{equation}
  {\rm CCH} + {\rm S} \rightarrow {\rm CCS} + {\rm H},
\end{equation}
and
\begin{subequations}
\begin{equation} 
{\rm S}^+ + {\rm CCH} \rightarrow {\rm CCS}^+ + {\rm H},
\end{equation}
\begin{equation} 
  {\rm CCS}^+ + {\rm H}_2 \rightarrow {\rm HCCS}^+ + {\rm H},
\end{equation}
\begin{equation} 
  {\rm HCCS}^+ + {\rm e} \rightarrow {\rm CCS} + {\rm H}.
\end{equation}
\end{subequations}

\noindent
In the reaction (7), the sulfur atom will attack the end carbon atom of CCH, which has the unpaired electron, to form the CCS structure (Yamada et al. 2002). In the case of the sulfur ion, the reaction may occur in a similar way. So C$^{13}$CS is made from $^{13}$CCH, and $^{13}$CCS is made from C$^{13}$CH through these processes. If these processes are efficient, the C$^{13}$CS/$^{13}$CCS ratio should be less than unity, given the C$^{13}$CH/$^{13}$CCH ratio is greater than unity. This is inconsistent with the observation: both the isotopomer ratios of CCH and CCS are greater than unity.
\subsubsection{Isotopomer-exchange reactions}
Fractionation by isotopomer-exchange reactions is also possible:
\begin{equation}
  {\rm ^{13}C}{\rm ^{12}CH} + {\rm H} \rightleftharpoons {\rm ^{12}C}{\rm ^{13}CH} + {\rm H} + \Delta E,
\end{equation}
\noindent
where $\Delta E$, the zero-point vibrational energy difference of $^{13}$CCH and C$^{13}$CH, is 8.1 K (R. Tarroni, private communication; Tarroni \& Carter 2003). As a candidate of the isotopomer-exchange reaction of CCS, on the other hand, Sakai et al. (2010) considered the following reaction:
\begin{equation}
  {\rm ^{13}C}{\rm ^{12}CS} + {\rm S} \rightleftharpoons {\rm ^{12}C}{\rm ^{13}CS} + {\rm S} + \Delta E,
\end{equation}  

\noindent
where $\Delta E$, the zero-point vibrational energy difference of $^{13}$CCS and C$^{13}$CS, is 17.4 K. According to the energy surface calculations, these exchange reactions, (9) and (10), do not have the activation barrier (Y. Osamura private communication). At low temperature ($T \sim$ 10 K), the isotopomer ratios of CCH and CCS can be greater than unity. 
 
\subsection{Reaction network}

We construct a chemical reaction network with carbon isotopes and isotopomers. The network is based on the gas-grain model developed by the Ohio State University astrochemistry group (Garrod \& Herbst 2006), and is expanded to include isotopes and isotopomers by the three procedures described below. While Garrod \& Herbst model consists of 655 species and 6309 reactions, our reaction network consists of 1203 gas and grain species, and 21116 gas phase and grain surface reactions. 

Firstly, we include $^{13}$C species into the reaction network statistically. For reactions involving $^{13}$C species, we assume that the total rate coefficient is the same as that of the corresponding reaction of the main isotope. We distinguish isotopomers only in species with less than three carbons to save computational time. For example, $^{13}$C$^{12}$C$^{12}$CH, $^{12}$C$^{13}$C$^{12}$CH and $^{12}$C$^{12}$C$^{13}$CH are different species, but $^{13}$C$^{12}$C$^{12}$C$^{12}$CH, $^{12}$C$^{13}$C$^{12}$C$^{12}$CH, $^{12}$C$^{12}$C$^{13}$C$^{12}$CH and $^{12}$C$^{12}$C$^{12}$C$^{13}$CH are treated as the same species, $^{12}$C$_3^{13}$CH. We checked molecular structures, so that symmetric molecules, such as C$_2$H$_2$, do not have isotopomers. We do not consider multiple $^{13}$C species for simplicity. The statistical expansion increases the number of the species and reactions by about two and three fold, respectively. For example, the reaction
\[
{\rm CCH} + {\rm O} \rightarrow {\rm CH} + {\rm CO},
\]
\noindent
with rate coefficients of $k$ is replaced by the following five reactions with rates $k$, $k/2$, $k/2$, $k/2$, and $k/2$, respectively:
\[  
{\rm ^{12}C}{\rm ^{12}CH} + {\rm O} \rightarrow {\rm ^{12}CH} + {\rm ^{12}CO},
\]
\[  
{\rm ^{12}C}{\rm ^{13}CH} + {\rm O} \rightarrow {\rm ^{13}CH} + {\rm ^{12}CO},
\]
\[  
{\rm ^{12}C}{\rm ^{13}CH} + {\rm O} \rightarrow {\rm ^{12}CH} + {\rm ^{13}CO},
\]
\[  
{\rm ^{13}C}{\rm ^{12}CH} + {\rm O} \rightarrow {\rm ^{13}CH} + {\rm ^{12}CO},
\]
\[  
{\rm ^{13}C}{\rm ^{12}CH} + {\rm O} \rightarrow {\rm ^{12}CH} + {\rm ^{13}CO}.
\] 

\noindent
In this way, statistical expansion is the main cause of increasing the number of species and reactions. "Statistically" means no fractionation occurs. In other words, all the isotope ratios are equal to the elemental [$^{12}$C/$^{13}$C] ratio and all the isotopomer ratios are unity.

Secondly, we include the gas phase reactions in Table 1 referring to the discussion in Section 2.2. The rate coefficients of these reactions are not measured in the laboratory. So we assume that the rate coefficients are 1 $\times$ 10$^{-10}$ cm$^3$ s$^{-1}$ for neutral-neutral reactions and 1 $\times$ 10$^{-9}$ cm$^3$ s$^{-1}$ for ion-neutral reactions. The dependence of our results on these rate coefficients is discussed in section 4.3.

Finally, we modify the branching ratios of proton transfer reactions and electron recombination reactions which break the C-H bond, such as
\[
{\rm H}{\rm ^{12}C}{\rm ^{12}CS}^+ + {\rm e} \rightarrow {\rm ^{12}C}{\rm ^{12}CS} + {\rm H},
\]
\[
{\rm H}{\rm ^{12}C}{\rm ^{13}CS}^+ + {\rm e} \rightarrow {\rm ^{12}C}{\rm ^{13}CS} + {\rm H},
\]
\[
{\rm H}{\rm ^{12}C}{\rm ^{13}CS}^+ + {\rm e} \rightarrow {\rm ^{13}C}{\rm ^{12}CS} + {\rm H}.
\]
\noindent   
In the statistically expanded network, the rate coefficients of these reactions are set to $k'$, $k'/2$, and $k'/2$, respectively, where $k'$ is the rate coefficient of the recombination of the $^{12}$C species. Then these reactions tend to make the isotopomer ratio of CCS unity, even if the isotopomer ratio of HCCS$^+$ are fractionated. But it would be more likely that these reactions do not rearrange the structure of C-C-S. So we set the rate coefficient to be $k'$, $k'$, and zero, respectively.

Table 2 lists the elemental abundances in our model; it is known as "low metal" values because of their strong depletions of species heavier than oxygen (e.g. Graedel et al. 1982). Initially the species are assumed to be atoms or atomic ions except for hydrogen, which is in molecular form. We assume the elemental isotope ratio, [$^{12}$C/$^{13}$C], of 60. If it is higher or lower than 60, the resultant molecular isotope ratios are scaled by [$^{12}$C/$^{13}$C]/60. We confirmed in the range of values, [$^{12}$C/$^{13}$C] = 30 - 77, the deviation from this scaling is less than 4 \%. 

We use the cosmic ray ionization rate of 1.3 $\times$ 10$^{-17}$ s$^{-1}$, which is commonly used for chemical model simulations for dense molecular clouds (e.g. Terzieva \& Herbst 1998). We assume that the sticking probability of colliding gaseous species onto grains is unity, and the adsorbed species can diffuse on the grain surfaces by thermal hopping. We adopt the same adsorption energies and non-thermal desorption rates as Garrod \& Herbst (2006). As a physical model, we assume a static cloud core. In our models, the physical parameters are set to $T$ = 10 K, $n_{\rm {H_2}}$ = 5 $\times$ 10$^4$ and $A_V$ = 10 mag, referring to the observation of TMC-1 (Pratap et al. 1997), unless otherwise stated. In section 3, we describe the results of our fiducial model. 

%% In this section, we use  the \subsection command to set off
%% a subsection.  \footnote is used to insert a footnote to the text.


%% This section also includes several of the displayed math environments
%% mentioned in the Author Guide.

\section{RESULT}
\subsection{Isotope ratios}
\subsubsection{Isotope ratios of gas species}

In this paper, we define 'abundance' as the fractional abundance with respect to H nuclei. Figure 1a shows the evolutions of major carbon-bearing species. Initially the carbon exists as carbon ion, which recombines to form carbon atoms in 10$^2$ yr. After a few 10$^4$ yr major form of carbon is mainly in carbon monoxide, which depletes onto the grain surfaces after 5 $\times$ 10$^4$ yr. 

Figure 1b shows the evolutions of the isotope ratios of the carbon-bearing species. Our results agree with Langer et al. (1984); carbon bearing species are divided into three groups: CO, HCO$^+$ and the "carbon isotope pool". Although we include gas-grain interactions, it does not affect the isotope ratios in the gas phase before the depletion of CO ($t <$ 10$^5$ yr). Adsorption rates are inversely proportional to square root of the mass of species, so that the rate of $^{12}$C species and $^{13}$C species are different. But this difference is small ($<$ 4 \%) and do not affect significantly. It should be noted, however, that when the CO depletion becomes significant, the reaction (1) becomes less efficient and isotope ratios of "carbon isotope pool" species are lowered compared to the model without gas-grain interactions. Since the thermal desorption is ineffective at 10 K, grain surface reactions do not affect isotope ratios of gaseous species. Then the isotope fractionation of the gaseous species is mostly determined by the gas phase reactions and the depletion of CO. Before a few 10$^4$ yr the abundance ratio of ${\rm CO/^{13}CO}$ is significantly smaller than 60 because of the reaction (1). As the abundance of CO increases with time and CO becomes the dominant carbon-bearing species, the isotope ratio gets close to the elemental abundance ratio of [$^{12}$C/$^{13}$C] = 60. For example, the ${\rm CO/^{13}CO}$ ratio is 54 at 10$^5$ yr. In contrast, the C$^+$/$^{13}$C$^+$ ratio is larger than 60 because of the reaction (1). Many carbon species are produced by reactions starting from C$^+$ or C, which is produced by recombination of C$^+$. Therefore the isotope ratios of these "carbon isotope pool" species are similar to the C$^+$/$^{13}$C$^+$ ratio before 10$^3$ yr and the C/$^{13}$C ratio after 10$^3$ yr, although actual values vary among species depending on their production pathways. For example, the CN/$^{13}$CN ratio is 83, while CS/$^{13}$CS ratio is 109 at 10$^4$ yr. The evolution of the HCO$^+$/H$^{13}$CO$^+$ ratio is similar to those of "carbon isotope pool" at $t <$ 10$^3$ yr, but the ratio is lowered by the reaction (2) after a few 10$^3$ yr; the HCO$^+$/H$^{13}$CO$^+$ ratio is 45 at 10$^5$ yr. 

To investigate the density dependence of the isotope ratios, we also performed calculations with $n_{\rm {H_2}}$ = 5 $\times$ 10$^3$ and 5 $\times$ 10$^5$ cm$^{-3}$. The evolution is in general faster at higher densities. Since we are interested in carbon-bearing species, we compare isotope ratios of these molecules at the time of peak carbon-chain abundance; 1 $\times$ 10$^5$, 2 $\times$ 10$^4$ and 4 $\times$ 10$^3$ yr for densities of $5 \times 10^3$, $5 \times 10^4$ and $5 \times 10^5$ cm$^{-3}$, respectively. We found that the isotope ratios do not significantly depend on density. For example, the ${\rm CN/^{13}CN}$ ratio at the selected time is 83, 81 and 80, and the ${\rm CS/^{13}CS}$ ratio is 83, 92 and 108 for densities of $5 \times 10^3$, $5 \times 10^4$ and $5 \times 10^5$ cm$^{-3}$, respectively. The isotope ratios of other species in each density model are listed in Table 3.   
 
\subsubsection{Isotope ratios of ice mantle species}

Figure 2(upper panels) shows the evolution of the major ice mantle species. H$_2$CO and CH$_3$OH are the most abundant organic species in ice mantles. They are formed from the carbon atom at $t < $ 2 $\times$ 10$^5$ yr and from CO at later time.
   
Figure 2(lower panels) shows the evolution of the isotope ratios of the ice mantle species. Carbon atoms and carbon monoxides in the gas phase are adsorbed and form more complex species by grain surface reactions. Before 10$^5$ yr, the isotope ratio of carbon atoms on grain surfaces is almost the same as in the gas phase. Carbon monoxide, on the other hand, is directly absorbed on grain surfaces and is also formed by the grain surface reactions of the carbon-bearing species. Then the isotope ratio of CO on grain surfaces is slightly different from that in the gas phase. The isotope ratios of ice mantle species depend mainly on whether the species are formed from the carbon atom or the CO molecule; the isotope ratio is larger than 60 if the species are formed from C atom, while the ratio is smaller than 60 if the species are formed from CO. For example, CH$_3$OH is formed from the carbon atom and its isotope ratio is 81 at $t$ = 2 $\times$ 10$^5$ yr. After 2 $\times$ 10$^5$ yr, CH$_3$OH are formed from the CO molecule and the isotope ratio decreases to 53 at 10$^6$ yr, which is almost the same as the isotope ratio of CO. HCOOH is an exception. It is formed from HCO$^+$ in the gas phase and adsorbed onto the grain surfaces. The isotope ratio of HCOOH is similar to that of HCO$^+$ in the gas phase. Isotope ratios of ice species in higher ($5 \times 10^5$ cm$^{-3}$) and lower ($5 \times 10^3$ cm$^{-3}$) density models are listed in Table 4.

Boogert et al. (2000; 2002) found that the solid CO$_2$/$^{13}$CO$_2$ ratio is similar to the solid CO/$^{13}$CO ratio towards NGC 7538 IRS 9, and suggested that CO$_2$ is formed from CO on grain surfaces, rather than from the carbon atom in the gas phase. In our model, CO$_2$ is formed in the gas-phase via the reaction of an O atom with HCO, which is formed from carbon atom, and adsorbed onto grain surfaces. Then the isotope ratio of CO$_2$ is similar to that of the carbon atom. However if we use the modified rates method (e.g. Ruffle and Herbst 2000), which artificially slows down the surface diffusion rate of the hydrogen atom, CO$_2$ is formed from CO on grain surfaces via the reaction of HCO + O (Aikawa et al. 2005). In this case, the isotope ratio of CO$_2$ ice at $t >$ 2 $\times 10^5$ yr should be closer to that of CO ice just like H$_2$CO and CH$_3$OH.

It should be noted that the observed isotope ratio of CO$_2$ ice varies with lines of sight. The solid CO/$^{13}$CO and CO$_2$/$^{13}$CO$_2$ ratios towards NGC 7538 IRS 9 are actually 71 $\pm$ 15 and 80 $\pm$ 11, respectively (Boogert et al. 2000; 2002). The isotope ratio of CO$_2$ is similar to or slightly higher than that of CO. On the other hand, towards NGC 2264 IRS1, Langer \& Penzias (1990) found the gas C$^{18}$O/$^{13}$C$^{18}$O ratio is 56 $\pm$ 3, and Gibb et al (2004) found the solid CO$_2$/$^{13}$CO$_2$ ratio is 131 $\pm$ 21. The isotope ratio of CO$_2$ ice is higher than that of CO gas by a factor of 2.3, assuming the carbon isotope fractionation of $^{18}$O species is similar to that of $^{16}$O species (e.g. Langer et al. 1984). In our model, in which CO$_2$ is mainly formed in the gas phase and adsorbed onto grain surfaces, such high isotope ratios of CO$_2$ ice relative to that of CO ice (gas) is realized at 10$^5$ yr in the model with $n_{\rm H_2}$ = $5 \times 10^3$ cm$^{-3}$. These variations may suggest that interstellar CO$_2$ ice is formed from both in the gas phase and on grain surfaces, and contributions of the two paths varies among clouds.

%Recently Geppert et al. (2010) found the isotope ratio of the gas methanol is almost similar to that of the gas CO. They also found the isotope ratio of the CO$_2$ is larger than them. 

%% The equation environment wil produce a numbered display equation.

%% The \notetoeditor{TEXT} command allows the author to communicate
%% information to the copy editor.  This information will appear as a
%% footnote on the printed copy for the manuscript style file.  Nothing will
%% appear on the printed copy if the preprint or
%% preprint2 style files are used.

%% The eqnarray environment produces multi-line display math. The end of
%% each line is marked with a \\. Lines will be numbered unless the \\
%% is preceded by a \nonumber command.
%% Alignment points are marked by ampersands (&). There should be two
%% ampersands (&) per line.

%% Putting eqnarrays or equations inside the mathletters environment groups
%% the enclosed equations by letter. For instance, the eqnarray below, instead
%% of being numbered, say, (4) and (5), would be numbered (4a) and (4b).
%% LaTeX the paper and look at the output to see the results.

%% This section contains more display math examples, including unnumbered
%% equations (displaymath environment). The last paragraph includes some
%% examples of in-line math featuring a couple of the AASTeX symbol macros.

\subsection{Isotopomer ratios}
\subsubsection{CCH}
 Figure 3a shows the temporal evolution of the assorted species relevant to the isotopomer fractionation of CCH and CCS. Figure 3b shows the C$^{13}$CH/$^{13}$CCH ratio. The isotopomer ratio is significantly enhanced in the fiducial model (solid line). In order to investigate whether the isotopomer fractionation of CCH is caused by (i) formation reactions or (ii) isotopomer-exchange reactions, we performed a calculation without the exchange reaction (9) (dashed line in Figure 3b). It is clear that isotopomer fractionation is mostly due to the exchange reaction.

Figure 4 shows the major formation and destruction paths of CCH isotopomers. The reactions (3) and (4) are the dominant formation reactions until 10$^4$ yr. The isotope ratio of CH$_2$ is larger than that of carbon atoms, and hence the isotopomer ratio is lowered. After 10$^4$ yr, on the other hand, CCH is mainly produced from the destruction of larger species (e.g. C$_3$H$_2^+$) by the oxygen atom and electron recombinations of symmetrical species (e.g. C$_2$H$_3^+$). These reactions tend to make the isotopomer ratio of CCH unity, since they form the same amount of the CCH isotopomer: C$^{13}$CH and $^{13}$CCH. Therefore, without the exchange reaction (9), the isotopomer ratio becomes almost unity in the model.

In our model, we do not distinguish isotopomers in species with more than four carbons. We checked if this simplification affects our results. We performed calculations without these large carbon chains (more than 4 carbon atom) and the exchange reaction (9), and confirmed the isotopomer ratio of CCH is as low as the dashed line in Figure 3b.

In the model with the exchange reaction, the isotopomer ratio increases with time in a few 10$^3$ yr as the electron abundance decreases. Then, the ratio gradually decreases as the H atom abundance decreases by adsorption and subsequent grain surface reactions. At several 10$^4$ yr, the ratio starts to increase again by the adsorption of oxygen atoms. After 3 $\times$ 10$^5$ yr, it reaches nearly the equilibrium ratio of 2.2 ($k_f$/$k_r$).

Since the H atom abundance is an important factor for the efficiency of the exchange reaction (9), we briefly discuss the formation and destruction of H atoms in dense cloud cores (Figure 3a). Before 10$^4$ yr, the primary sources of the H atom are the reactions of hydrocarbons, such as the recombination of CH$_2^+$ and the reaction of CH + O, and the destruction is by adsorption on grain surfaces followed by subsequent hydrogenation reactions, such as H + OH. At $t > 10^{4}$ yr, H atoms are mainly formed by cosmic ray induced photodissociation of hydrogen molecules. At this stage the absolute abundance of the H atom is $\sim$1 cm$^{-3}$ independent of the gas density. We also checked the dependence of our results on the initial H atom abundance; if the initial abundance is lowered by an order of magnitude, the isotopomer ratio varies only slightly. 

Figure 5 shows density dependence of the C$^{13}$CH/$^{13}$CCH ratio. As the density increases, both the abundances of electron and H atom decrease. While the electron abundance is proportional to $n_{\rm H_2}^{-0.5}$, the H atom abundance is proportional to $n_{\rm H_2}^{-1}$. Then the peak isotopomer ratio decreases with density. 

The isotopomer fractionation also affects the isotope ratios, i.e. the CCH/$^{13}$CCH and CCH/C$^{13}$CH ratios (Figure 6). The reaction (9) enhances $^{13}$C in C$^{13}$CH, while it dilutes $^{13}$C in $^{13}$CCH. For comparison, the dashed line in Figure 6 shows the isotope ratios of CCH in the model without the isotopomer fractionation: the model without the reaction (3), (4), (5), (6), (9) and (10).
\subsubsection{CCS}

Figure 7 shows the temporal evolution of the C$^{13}$CS/$^{13}$CCS ratio. The isotopomer ratio of CCS is lower than unity in the model without the exchange reaction (10) (dashed line). The reactions (5) and (6) are not the main formation reactions of C$^{13}$CS and $^{13}$CCS in our model. Instead, CCS is mainly formed by the electron recombination of larger ions, such as ${\rm HC_2S^+}$ (Figure 4). These reactions tend to make the isotopomer ratio of CCS unity. The reaction (7) and (8) decreases the isotopomer ratio of CCS, because the isotopomer ratio of CCH is larger than unity. In the fiducial model (solid line), the isotopomer ratio is lower than unity except for very limited periods. The isotopomer ratio reaches a maximum of about 1.2 by the reaction (10) at 3 $\times$ 10$^4$ yr. Figure 8 shows the density dependence of the C$^{13}$CS/$^{13}$CCS ratio. Although the peak isotopomer ratio increases with density, the maximum ratio is 1.8, which is much lower than observed. Unlike CCH, the exchange reaction (10) is inefficient, because the sulfur atom is less abundant than the hydrogen atom by three orders of magnitude.

\section{DISCUSSION}
\subsection{Comparisons with observations of TMC-1}
The gas density in TMC-1 is 10$^4$ - 10$^5$ cm$^{-3}$ (Pratap et al. 1997). In the model with $n_{\rm {H_2}}$ = 5 $\times$ 10$^4$ cm$^{-3}$, the abundances of carbon-chain molecules, such as HC$_3$N, reach their maxima at 2 $\times$ 10$^4$ yr. Since carbon-chain molecules are abundant in TMC-1, we compare the obtained molecular abundaces and isotope and isotopomer ratios in the fiducial model at 2 $\times$ 10$^4$ yr to the observed values in TMC-1.

Pratap et al. (1997) made maps along the TMC-1 ridge, and determine the column densities of C$^{18}$O and CCH. The molecular abundance of CCH in TMC-1 is 5.7 $\times$ 10$^{-10}$ - 4.0 $\times$ 10$^{-9}$, assuming a C$^{18}$O/H$_2$ ratio of 1.7 $\times$ 10$^{-7}$ (Frerking et al. 1982). In our model, we obtain the abundance of CCH to be 3.0 $\times$ 10$^{-9}$, which is in reasonable agreement with the observed value.     

In the fiducial model, we obtain the CCH/$^{13}$CCH, CCH/C$^{13}$CH and C$^{13}$CH/$^{13}$CCH ratios to be 99, 63 and 1.6, respectively, at 2 $\times$ 10$^4$ yr. The observed CCH/C$^{13}$CH and CCH/$^{13}$CCH ratios are higher than 170 and 250, respectively (Sakai et al. 2010). The average isotope ratio, which is calculated by
\[
2\times \frac{{\rm CCH}}{{\rm C^{13}CH}+{\rm ^{13}CCH}},
\]
\noindent
is higher than 200, and is three times higher than our model. Additional isotope-exchange reactions, which are currently unknown, may be needed to reproduce the observed values. On the other hand, the observed C$^{13}$CH/$^{13}$CCH ratio is 1.6 $\pm$ 0.4 (Sakai et al. 2010), which is in reasonable agreement with our model. Therefore the exchange reaction (9) can account for the isotopomer fractionation of CCH.  

We obtain the CCS/$^{13}$CCS, CCS/C$^{13}$CS and C$^{13}$CS/$^{13}$CCS ratios to be 84, 76 and 1.1, respectively, in the fiducial model at 2 $\times$ 10$^4$ yr. The observed CCS/$^{13}$CCS, CCS/C$^{13}$CS and C$^{13}$CS/$^{13}$CCS ratios are 230 $\pm$ 130, 54 $\pm$ 5, and 4.2 $\pm$ 2.3, respectively (Sakai et al. 2007). The average isotope ratio is 87, which is in good agreement with our model results. But our model underestimates the isotopomer fractionation. We discuss this problem in the next section. 

\subsection{Exchange reactions of CCS}
The isotopomer-exchange reaction (10) is not efficient enough to account for the observed isotopomer ratio of CCS, because of the low abundance of sulfur atoms. Since the reactant is not necessarily the sulfur atom for the exchange reaction, we consider a possibility of the following neutral-neutral reaction:
\begin{equation}
  ^{13}{\rm C}^{12}{\rm CS} + {\rm H} \rightleftharpoons {\rm ^{12}C^{13}CS} + {\rm H} + \Delta E,
\end{equation}
\noindent
where $\Delta E$ is 17.4 K. This can be regarded as a catalytic reaction by the hydrogen atom. According to the quantum chemical calculation by Yamada et al. (2002), the reaction (11) does not have the activation barrier. It could be more effective than the reaction (10), because the hydrogen atom is more abundant than the sulfur atom. Since the rate coefficients of the reaction (11) has not been measured or calculated, we simply assume that the forward rate coefficient is 1 $\times$ 10$^{-10}$ cm$^3$ s$^{-1}$, which is a typical collisional rate coefficient between neutral species. Figure 9a shows the temporal evolution of the isotopomer ratio of CCS in the model with the exchange reaction (11). The ratio is significantly enhanced in comparison with the model in Figures 7 and 8, and reaches nearly the equilibrium ratio of 5.7 ($k_f$/$k_r$) at a few 10$^4$ yr. After several 10$^4$ yr, CCS abundance decreases by about two orders of magnitude (Figure 3a). The abundance of S-bearing molecular ions, such as HCCS$^+$, decrease only by an order of magnitude, because H$_3^+$ abundance is enhanced by the CO depletion. Then the exchange reaction (11), whose efficiency is proportional to the CCS abundance, becomes less effective than ion-molecule reactions, and the isotopomer ratio decreases to 4.1. Our model with the reaction (11) can reasonably explain the observed isotopomer ratio. 

The isotopomer fractionation also affects the isotope ratios, CCS/$^{13}$CCS and CCS/C$^{13}$CS (Figure 9b). $^{13}$C is significantly diluted in $^{13}$CCS, and the CCS/$^{13}$CCS ratio is higher than 200 in several 10$^4$ yr. On the other hand, the CCS/C$^{13}$CS ratio is lower than 60. At 2 $\times$ 10$^4$ yr, the CCS/$^{13}$CCS and CCS/C$^{13}$CS ratios are 263 and 47, respectively, which are in reasonable agreement with the observed isotope ratios. Hence the reaction (11) is a possible culprit for the isotopomer fractionation of CCS.

\subsection{Dependences on rate coefficients of isotopomer-exchange reactions}
In the previous sections we assumed the forward reaction rate coefficients of the isotopomer-exchange reactions are all 1 $\times$ 10$^{-10}$ cm$^3$ s$^{-1}$, which is a typical collisional rate coefficient between neutral species. However it is not obvious whether the actual exchange reactions proceed at the collisional rate. For example, the rate coefficient of the similar type of reactions, such as HNC$_3$ + H $\rightarrow$ HC$_3$N + H, is assumed to be 1 $\times$ 10$^{-11}$ cm$^{-3}$ s$^{-1}$ in our model (e.g. Garrod $\&$ Herbst 2006). In particular, reaction (11) requires the rearrangement of the order of the carbon nuclei in the chain through the transient cyclic isomers (Yamada et al. 2002). Although the reaction is exothermic, the efficiency of reaction (11) might be low. Hence, we investigate the dependence of the isotopomer ratios on the rate coefficients of reactions (9) and (11). Figure 10 shows the temporal evolution of the isotopomer ratios of (a) CCH and (b) CCS. The forward reaction rate coefficients are 1 $\times$ 10$^{-10}$ cm$^3$ s$^{-1}$ (solid line), 1 $\times$ 10$^{-11}$ cm$^3$ s$^{-1}$ (dashed line) and 1 $\times$ 10$^{-12}$ cm$^3$ s$^{-1}$ (dotted line). We can see the forward reaction rate coefficients should be larger than 1 $\times$ 10$^{-11}$ cm$^3$ s$^{-1}$ to account for the observed isotopomer ratios by the exchange reactions (9) and (11). Laboratory measurements, as well as more detailed theoretical calculations of these isotopomer-exchange reactions, are highly desirable.

%The C$^{13}$CH/$^{13}$CCH ratio at 2 $\times$ 10$^4$ yr is 1.1, 1.0 for $k_f$ of 10$^{-11}$, 10$^{-12}$ cm$^3$s$^{-1}$, respectively. The C$^{13}$CS/$^{13}$CCS ratio at 2 $\times$ 10$^4$ yr is 4.0, 2.5 for $k_f$ of 10$^{-11}$, 10$^{-12}$ cm$^3$s$^{-1}$, respectively.

\subsection{Initial condition of the carbon}
So far, we have assumed that all the carbon initially exists as carbon ions. This assumption is the most favorable for the carbon isotope fractionation, because the isotope-exchange reaction (1) becomes efficient. Here, we investigate the dependencies of isotope ratios on the initial condition of the carbon.

We performed calculations with two different initial conditions; (1) a half of the carbon exists as atomic carbon, and (2) all the carbon is in atomic form. In the former case, the evolution of isotope ratios in the gas phase is very similar to the fiducial model, because the abundances of major carbon-bearing species are almost the same as the fiducial model after 10$^2$ yr. Figure 11 shows the evolutions of (a) molecular abundances and (b) isotope ratios in the gas phase for the latter case: all the carbon is initially atomic carbon. The molecular abundances and isotope ratios are different from the fiducial model before 10$^5$ yr, but they approach to the fiducial model after a few 10$^5$ yr. The abundances of C, C$^+$ and CO are similar to those in the fiducial model after 10$^4$ yr. The isotope ratio of CO is nearly 60 before 10$^4$ yr, since the abundance of C$^+$ is small. After 10$^4$ yr the ratio slightly decreases: $^{12}$CO/$^{13}$CO = 56 at 10$^6$ yr. On the other hand, the isotope ratio of C$^+$ is as high ($>$ 60) as in the fiducial model, since the abundance of CO is smaller than in the fiducial model only by an order of magnitude. The isotope ratio of the carbon atom is nearly 60 before 10$^5$ yr. After 10$^5$ yr the ratio becomes greater than 60 because of the neutralization of the carbon ion. "Carbon isotope pool" species are mostly produced from the carbon atoms, then their isotope ratios are similar to that of the carbon atom. In this way, the carbon isotope fractionation occurs by the isotope-exchange reactions after 10$^5$ yr, even if all the carbon is initially in the carbon atom. These results suggest isotope fractionation in early times does not much affect the isotope ratios after 10$^5$ yr. In order to account for the isotope fractionation at the period of peak carbon chain abundances ($t \sim 2\times 10^4$ yr), however, some fraction of carbon should be carbon ion initially.
%In the case of (b), which is $^{13}$C-rich case, however, the deviation from linearity is as large as 10 \%. The reason is that we do not consider multiple $^{13}$C species; for example we ignore the following reaction,
%when we expand the network statistically. The reaction (12) is 6 times less frequently than reactions including single $^{13}$C species.  
%\subsection{Effects of photo dissociation reactions to isotopomer ratios}
%In the diffuse clouds which are precursors of cloud cores, photo dissociation reactions might affect the isotopomer ratios.  

%In optically thin regions, isotope fractionation occurs by photo dissociation reactions. We assume optically thick (A$_{\rm V}$ = 10) quasi-static cloud cores as a physical model, then we ignore this effect.    

\subsection{Summary}
We constructed a gas-grain chemical network model that includes carbon isotopes and isotopomers in order to investigate the evolution of molecular abundances, the carbon isotope ratios and the isotopomer ratios of CCH and CCS in cold dense cloud cores. The principal results are follows.

1. We confirm that the isotope ratios of molecules, both in the gas phase and on grain surfaces, mostly depend on whether the species is formed from the carbon atom (ion) or the CO molecule; the isotope ratio is larger than the elemental abundance ratio of [$^{12}$C/$^{13}$C] if the species is formed from the carbon atom, while the ratio is smaller if the species is formed from the CO molecule.

2. There are two possible processes of isotopomer fractionation: (i) the formation pathways of the species and (ii) the exchange of the $^{13}$C position by isotopomer-exchange reactions. We reproduce the observed isotopomer ratio C$^{13}$CH/$^{13}$CCH by considering the isotopomer-exchange reaction: $^{13}$CCH + H $\rightarrow$ C$^{13}$CH + H + 8.1 K, if this reaction proceeds with the forward rate coefficient  $k_f$ $>$ 10$^{-11}$ cm$^3$ s$^{-1}$. The isotope ratio is underestimated in our model, which is left for future studies.

3. We reproduce the observed C$^{13}$CS/$^{13}$CCS, CCS/C$^{13}$CS and CCS/$^{13}$CCS ratios simultaneously by considering the isotopomer-exchange reaction: $^{13}$CCS + H $\rightarrow$ C$^{13}$CS + H + 17.4 K, if this reaction proceeds with the forward rate coefficient  $k_f$ $>$ 10$^{-11}$ cm$^3$ s$^{-1}$. 

4. In conclusion, isotopomer fractionation of CCH and CCS can be due to isotopomer-exchange reactions. Laboratory measurements and detailed quantum calculations of these isotopomer exchange reactions are highly desired.\\

%% The displaymath environment will produce the same sort of equation as
%% the equation environment, except that the equation will not be numbered
%% by LaTeX.


%% Included in this acknowledgments section are examples of the
%% AASTeX hypertext markup commands. Use \url without the optional [HREF]
%% argument when you want to print the url directly in the text. Otherwise,
%% use either \url or \anchor, with the HREF as the first argument and the
%% text to be printed in the second.

\acknowledgments
Acknowledgements. We thank Yoshihiro Osamura for providing the zero-point energies of C$^{13}$CS and $^{13}$CCS, and valuable discussions. We also thank Riccardo Tarroni for providing the zero-point energies of C$^{13}$CH and $^{13}$CCH. We would like to thank the anonymous referee for the helpful comments to improve the manuscript. This work is supported by a grant-in-aid for scientific research (21244021) and Global COE program "Foundation of International Center for Planetary Science" (G11) of the Ministry of Education, Culture, Sports, Science and Technology of Japan (MEXT). 

\clearpage

%% the file, uncomment the \plotone, \plottwo, and
%% If you need a layout that cannot be achieved with \plotone or
%% \plottwo, you can invoke the graphicx package directly with the
%% \includegraphics command or use \plotfiddle. For more information,
%% please see the tutorial on "Using Electronic Art with AASTeX" in the
%% documentation section at the AASTeX Web site,
%% http://www.journals.uchicago.edu/AAS/AASTeX.
%%
%% The examples below also include sample markup for submission of
%% supplemental electronic materials. As always, be sure to check
%% the instructions to authors for the journal you are submitting to
%% for specific submissions guidelines as they vary from

%% This example uses \plotone to include an EPS file scaled to
%% 80% of its natural size with \epsscale. Its caption
%% has been written to indicate that additional figure parts will be
%% available in the electronic journal.

\begin{figure}
\epsscale{.70}
\plotone{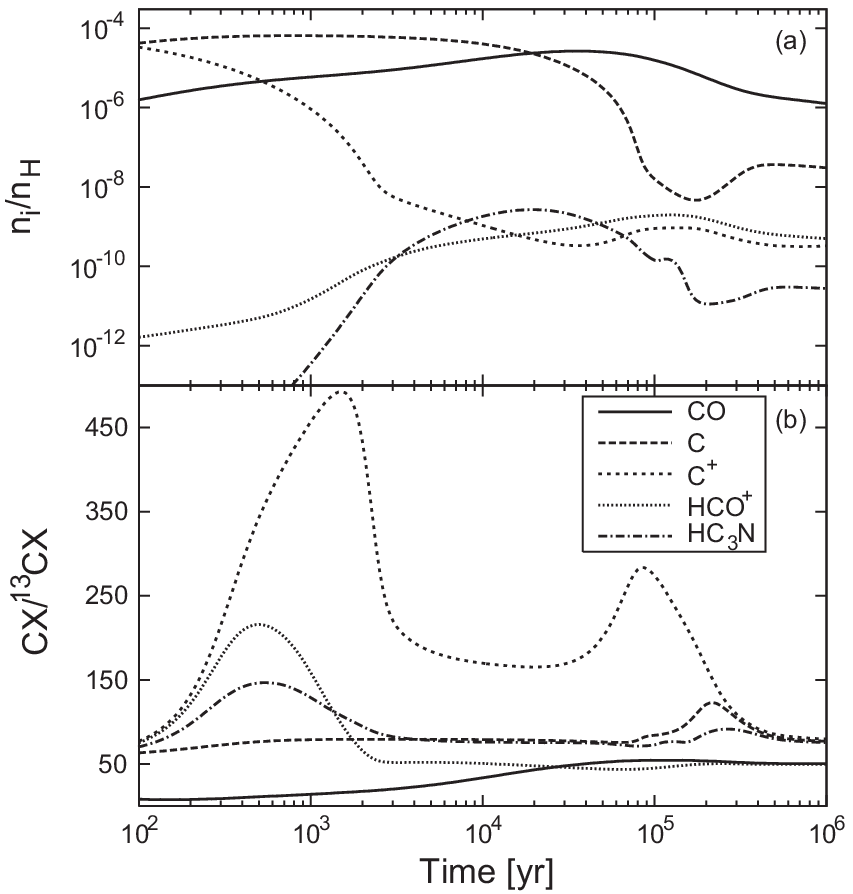}
\caption{Temporal variation of (a) molecular abundances and (b) isotope ratios in the gas phase in the fiducial model.\label{fig1}}
\end{figure}

\clearpage

\begin{figure}
\epsscale{1.0}
\plotone{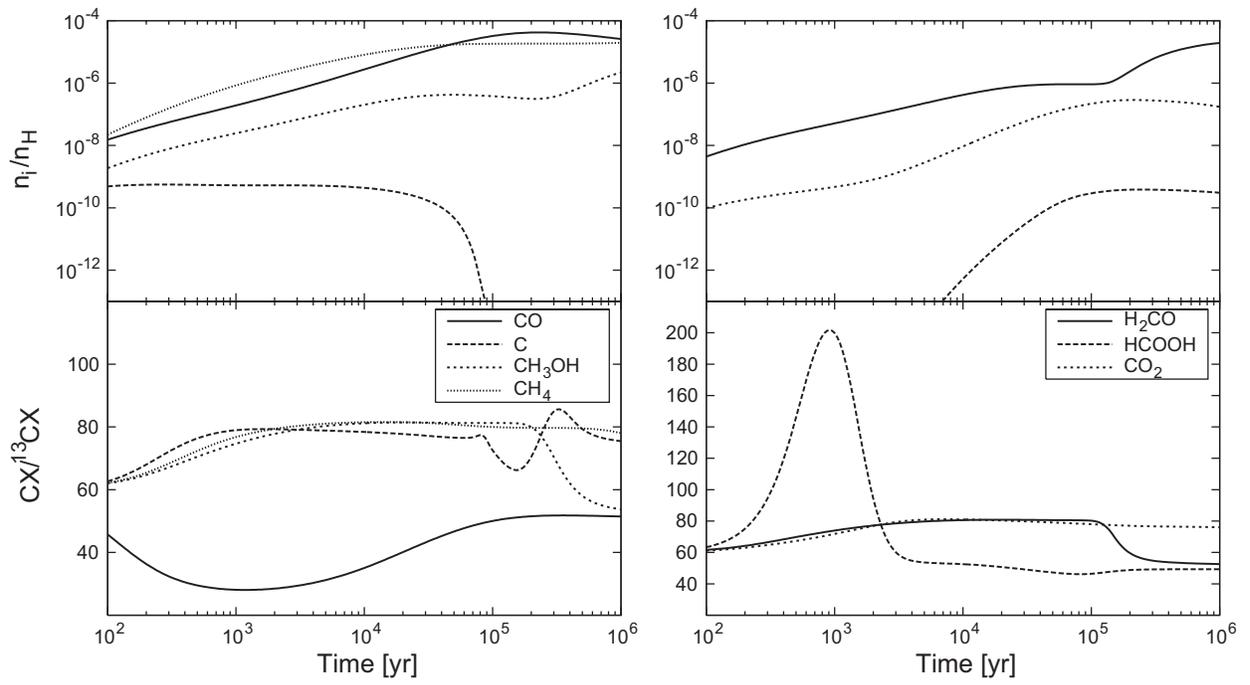}
\caption{Temporal variation of molecular abundances (upper panels) and isotope ratios (lower panels) of ice mantle species in the fiducial model.\label{fig2}}
\end{figure}

\begin{figure}
\epsscale{.70}
\plotone{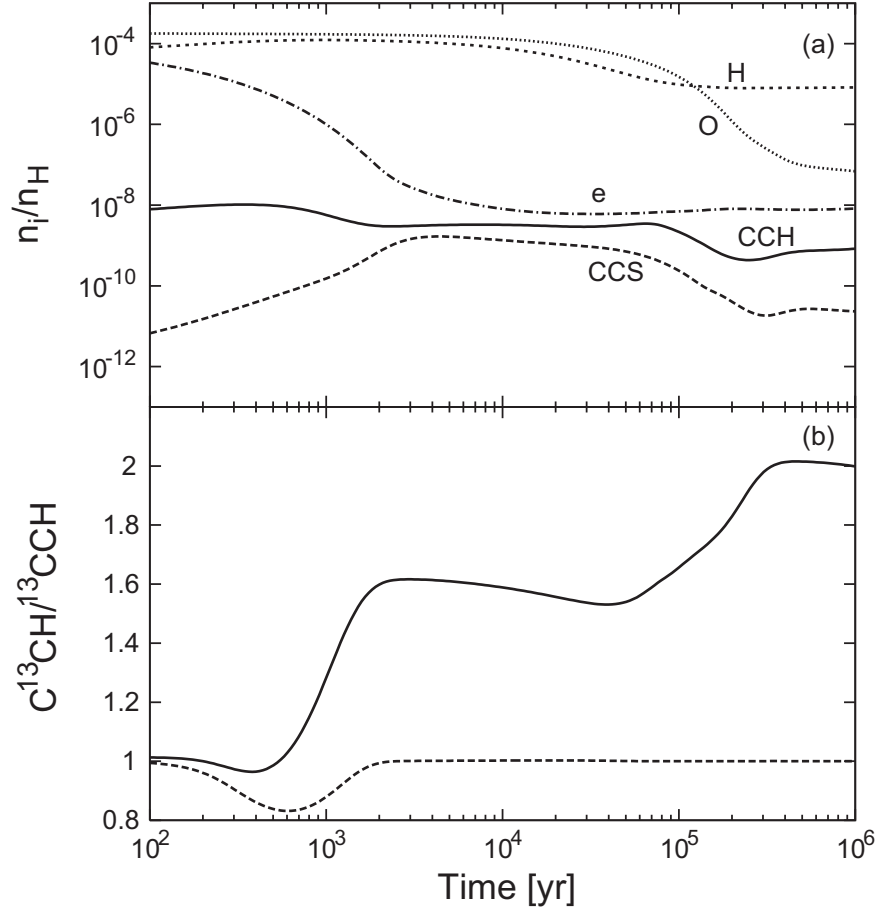}
\caption{(a) Temporal variation of the abundances of CCH, CCS, H atom, O atom and electron. (b) Temporal variation of the C$^{13}$CH/$^{13}$CCH ratio in the fiducial model (solid line) and in the model without the exchange reaction (8) (dashed line).\label{fig3}}
\end{figure}

\begin{figure}
\epsscale{.70}
\plotone{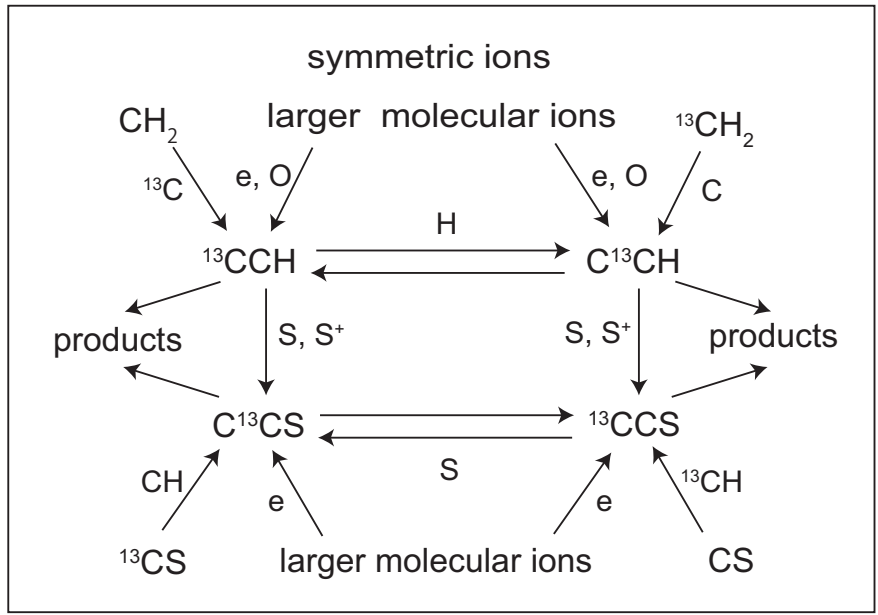}
\caption{Production and destruction pathways of CCH and CCS.\label{fig4}}
\end{figure}

\begin{figure}
\epsscale{.70}
\plotone{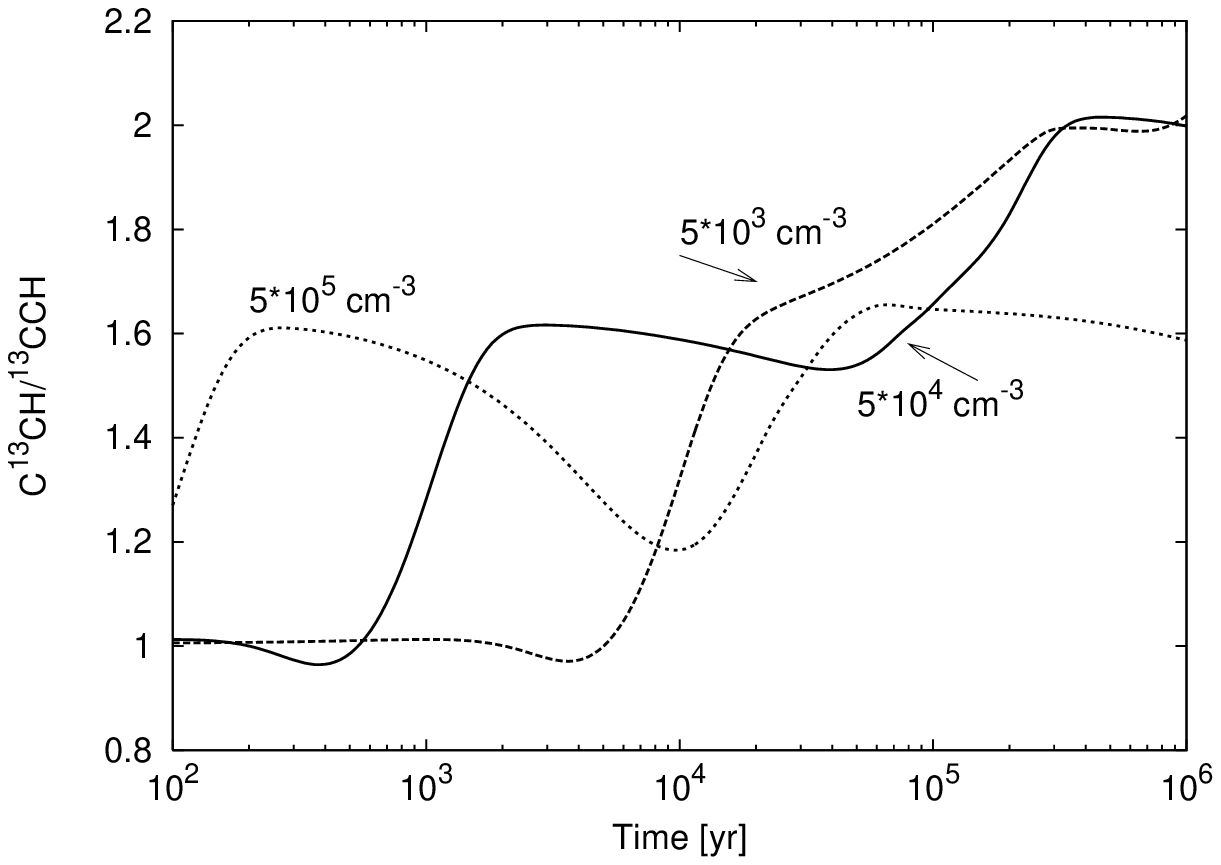}
\caption{Isotopomer ratio of CCH in the model with $n_{\rm H_2}$ = 5$\times$10$^{3}$ (dashed line), 5$\times$10$^{4}$ (solid line), and 5$\times$10$^{5}$ cm$^{-3}$ (dotted line).\label{fig5}}
\end{figure}

\begin{figure}
\epsscale{.70}
\plotone{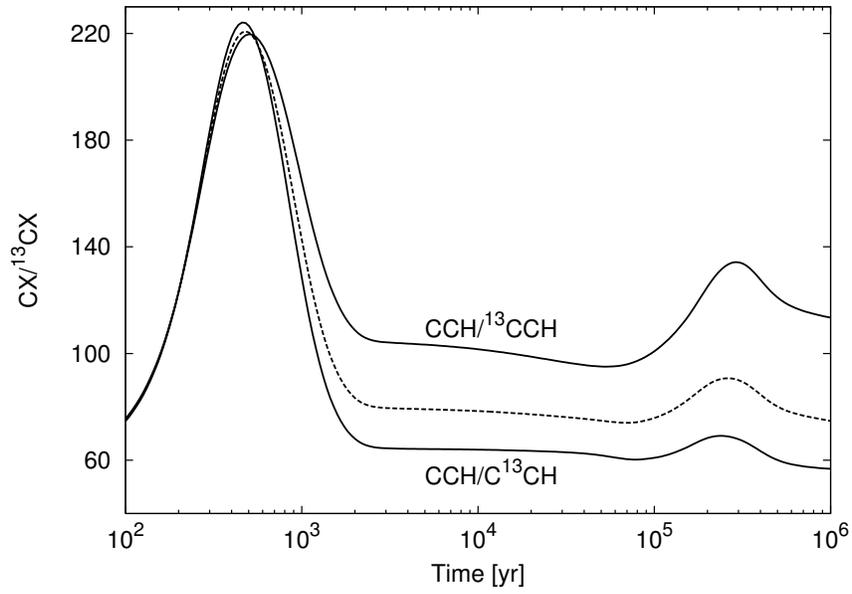}
\caption{Temporal variation of isotope ratios of CCH: the CCH/$^{13}$CCH and CCH/C$^{13}$CH ratios. The solid lines show ratios in the fiducial model, whereas the dashed line shows the ratio in the model without isotopomer fractionation.\label{fig6}}
\end{figure}

\begin{figure}
\epsscale{.70}
\plotone{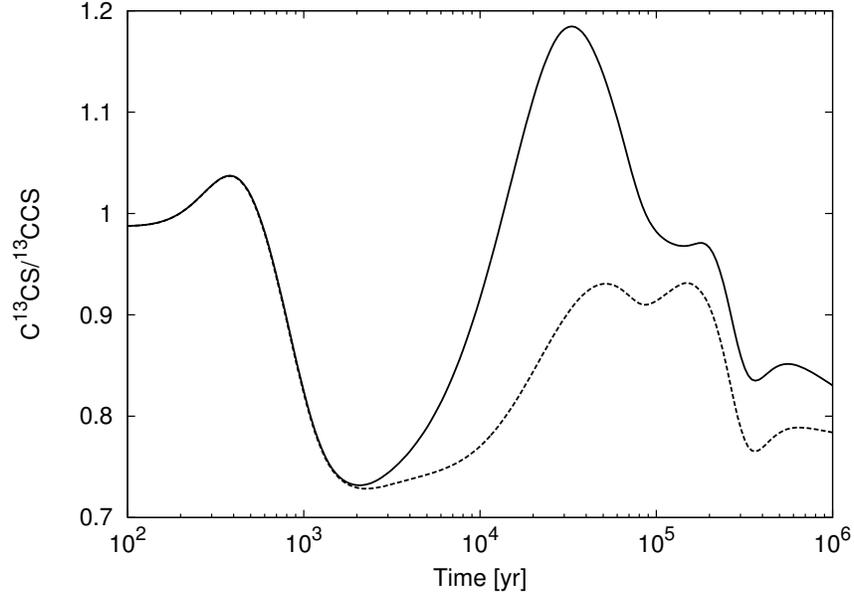}
\caption{Temporal variation of the C$^{13}$CS/$^{13}$CCS ratio in the fiducial model (solid line) and in the model without reaction (9) (dashed line).\label{fig7}}
\end{figure}

\begin{figure}
\epsscale{.70}
\plotone{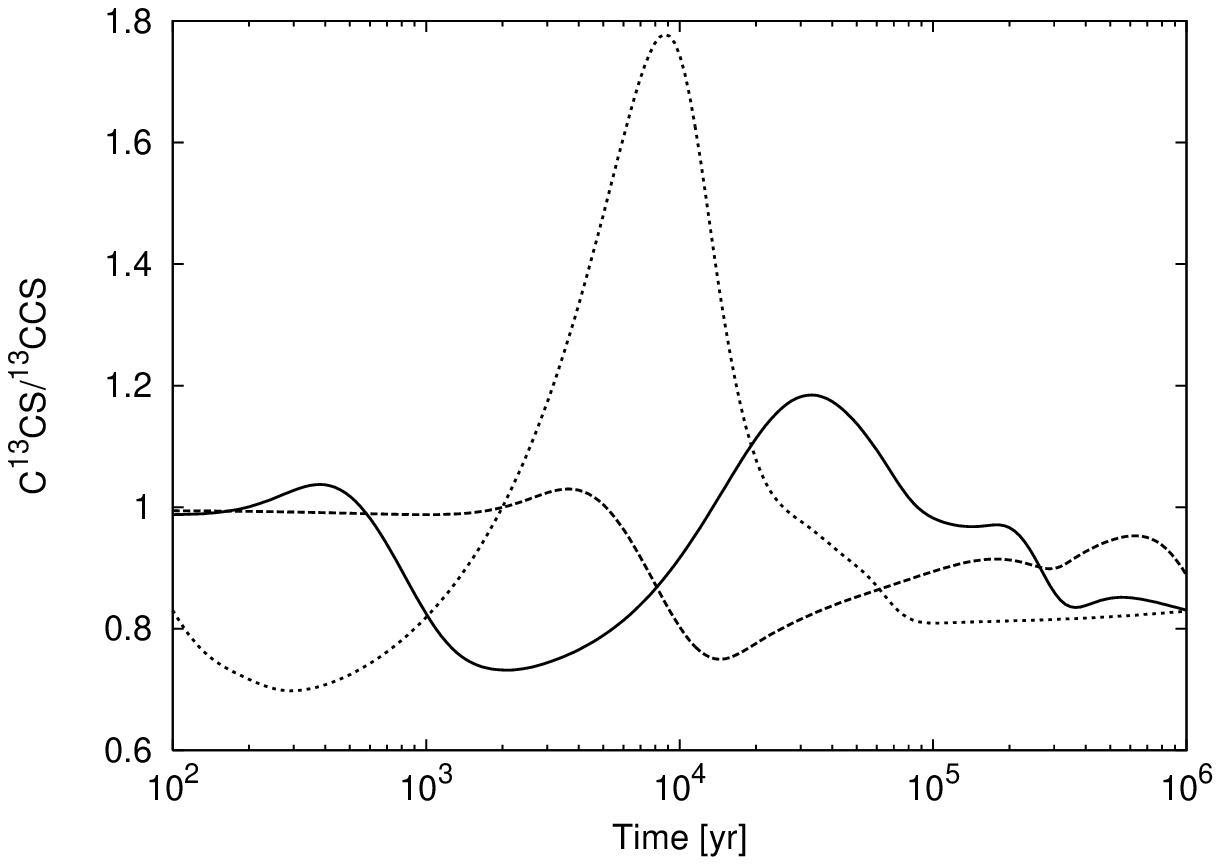}
\caption{Isotopomer ratio of CCS in the model with $n_{\rm H_2}$ = 5$\times$10$^{3}$ (dashed line), 5$\times$10$^{4}$ (solid line), and 5$\times$10$^{5}$ cm$^{-3}$ (dotted line).\label{fig8}}
\end{figure}

%\begin{figure}
%\epsscale{.70}
%\plotone{fig9.eps}
%\caption{Temporal variation of the C$^{13}$CS/$^{13}$CCS ratio in the model with reaction (13), (solid line) and in the standard model, (dashed %line). The density is n$_{\rm H_2}$ = 5$\times$10$^{4}$ cm$^{-3}$\label{fig9}}
%\end{figure}

\begin{figure}
\epsscale{1.0}
\plotone{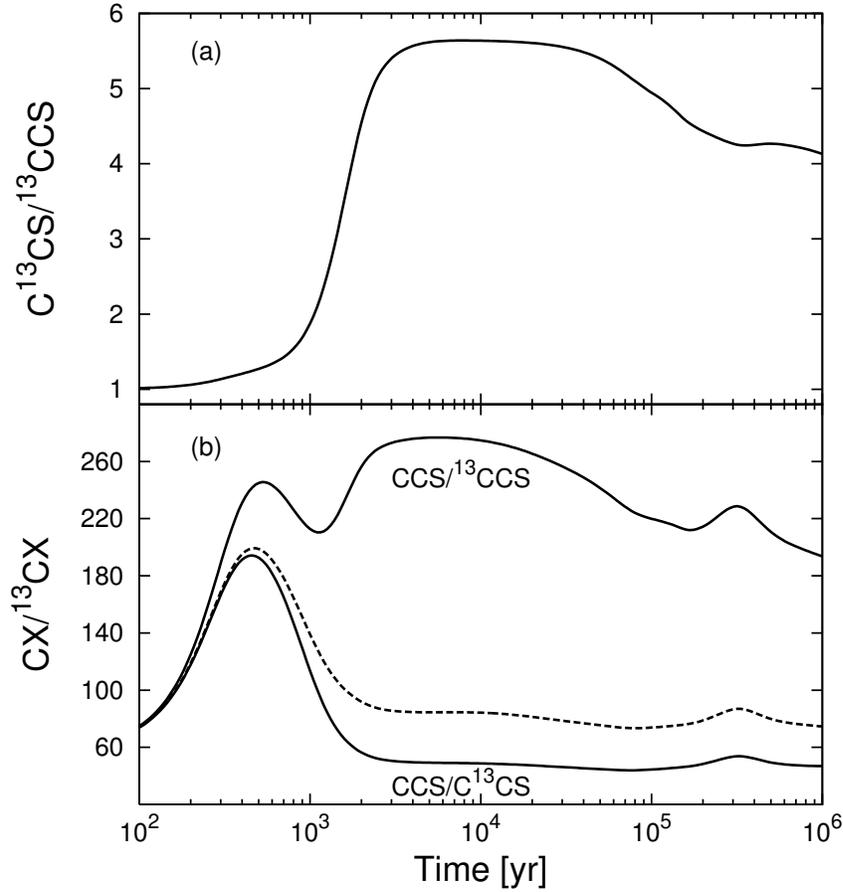}
\caption{Temporal variation of (a) isotopomer ratio, C$^{13}$CS/$^{13}$CCS and (b) isotope ratios of CCS. The solid lines show the ratios in the fiducial model, while the dashed line shows the isotope ratio of CCS in the model without the isotopomer fractionation.\label{fig9}}
\end{figure}

\begin{figure}
\epsscale{1.0}
\plotone{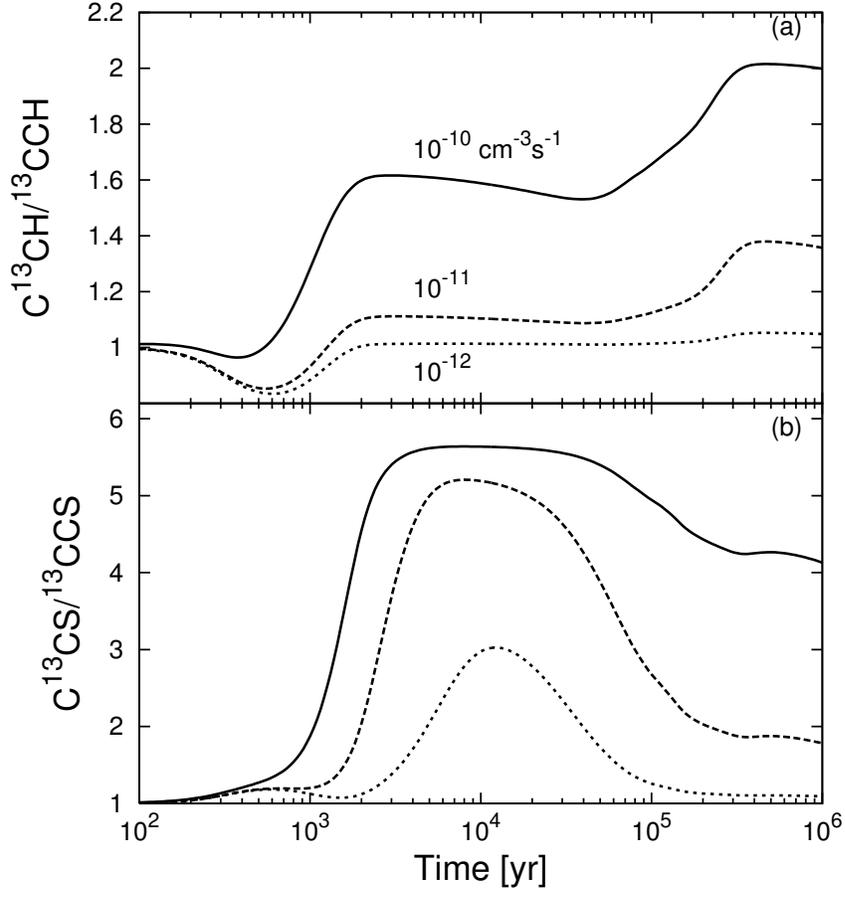}
\caption{Temporal variation of the (a) C$^{13}$CH/$^{13}$CCH ratio, and (b) C$^{13}$CS/$^{13}$CCS ratio. Forward rate coefficients of the reaction (9) and (13) are 1 $\times$ 10$^{-10}$ cm$^3$s$^{-1}$ (solid line), 1 $\times$ 10$^{-11}$ (dashed line), and 1 $\times$ 10$^{-12}$ cm$^3$s$^{-1}$ (dotted line).\label{fig10}}
\end{figure}

\begin{figure}
\epsscale{1.0}
\plotone{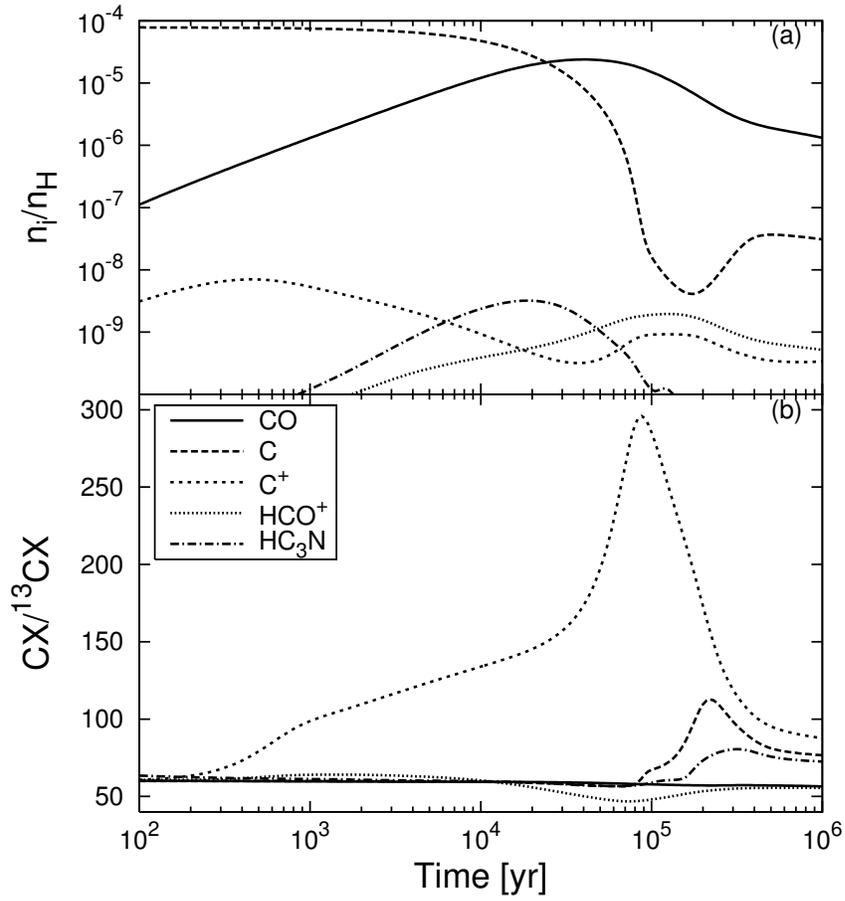}
\caption{Same as Figure 1, but all the carbon is in carbon atoms initially.\label{fig11}}
\end{figure}

%\begin{figure}
%\epsscale{1.0}
%\plotone{fig13.eps}
%\caption{Dependence of isotope ratios on the elemental [$^{12}$C/$^{13}$C] ratio: (a) [$^{12}$C/$^{13}$C] = 77 and (b) [$^{12}$C/$^{13}$C] = 6. $R$ is defined by $ \left(\frac{^{12}{\rm CX}}{^{13}{\rm CX}}\right)_{60} / \left(\frac{^{12}{\rm CX}}{^{13}{\rm CX}}\right)_{x} \times \frac{x}{60}$, $x$ is the value of the [$^{12}$C/$^{13}$C] ratio. The density is n$_{\rm H_2}$ = 5$\times$10$^{4}$ cm$^{-3}$.\label{fig11}}
%\end{figure}
%\begin{figure}
%\epsscale{.70}
%\plotone{discussion_ccs_isotope.eps}
%\caption{Calculated isotope ratio of CCS versus time in the case of considering the reaction; $^{13}$CCS + H $\rightleftharpoons$ C$^{13}$CS + H + 15K. Solid:k=10$^{-10}$cm$^3$s$^{-1}$, Dash:k=10$^{-11}$cm$^3$s$^{-1}$, Dot:k=0.\label{fig9}}
%\end{figure}

\clearpage

\renewcommand\thefootnote{\alph{footnote}} 

\begin{table}
\begin{center}
\caption{Rate coefficients and zero point energy differences of the reactions added to the statistically expanded network}
\begin{tabular}{lcc}
\tableline\tableline
\hspace{60pt}Reaction &$k$ (cm$^3$ s$^{-1}$) \footnotemark &$\Delta E$ $(K)$\\
\tableline
CH +  CS $\rightarrow$ CCS + H &1.00(-10) \footnotemark&-\\
CH +  $^{13}$CS $\rightarrow$ C$^{13}$CS + H &1.00(-10)&-\\
$^{13}$CH + CS $\rightarrow$  $^{13}$CCS + H &1.00(-10)&-\\
CCS$^+$ + H$_2$ $\rightarrow$ HCCS$^+$ + H &1.00(-9)&-\\
C$^{13}$CS$^+$ + H$_2$ $\rightarrow$ HC$^{13}$CS$^+$ + H &1.00(-9)&-\\
$^{13}$CCS$^+$ + H$_2$ $\rightarrow$ H$^{13}$CCS$^+$ + H &1.00(-9)&-\\
$^{13}$C$^{+}$ + CO $\rightleftharpoons$ $^{13}$CO + C$^{+}$ &1.34(-9) &35\\
HCO$^{+}$ + $^{13}$CO $\rightleftharpoons$ CO + H$^{13}$CO$^{+}$ &6.50(-10) &9\\
$^{13}$CCH + H $\rightleftharpoons$ C$^{13}$CH + H &1.00(-10) &8.1\\
$^{13}$CCS + S $\rightleftharpoons$ C$^{13}$CS + S &1.00(-10) &17.4\\
\tableline
\end{tabular}
\tablecomments{(a) For exchange reactions $k$ are the rate coefficients of forward reactions. (b) $x(-y)$ represents $x\times 10^{-y}$.}
%% Any table notes must follow the \end{tabular} command.
\end{center}
\end{table}

%\clearpage
%% Any table notes must follow the \end{tabular} command.

%% Text for table notes should follow after the \enddata but before
%% the \end{deluxetable}. Make sure there is at least one \tablenotemark
%% in the table for each \tablenotetext.

%% If you use the table environment, please indicate horizontal rules using
%% Do not put multiple tabular environments within a single table.
%% The optional \label should appear inside the \caption command.

\clearpage

\begin{table}
\begin{center}
\caption{Initial abundances with respect to hydrogen nuclei}
\begin{tabular}{cc}
\tableline\tableline
species & Abundance \\
\tableline
H$_2$.............................................. &5.00(-1)\\
H............................................... &5.00(-5)\\
He.............................................. &9.75(-1)\\
$^{12}$C$^+$............................................ &7.86(-5)\\
$^{13}$C$^+$............................................ &1.31(-6)\\
N............................................... &2.47(-5)\\
O............................................... &1.80(-4)\\
Si$^+$............................................. &2.74(-9)\\
S$^+$.............................................. &9.14(-8)\\
Fe$^+$............................................. &2.74(-9)\\
Na$^+$............................................. &2.25(-9)\\
Mg$^+$............................................. &1.09(-8)\\
Cl$^+$............................................. &1.00(-9)\\
P$^+$.............................................. &2.16(-10)\\
\tableline
\end{tabular}
%% Any table notes must follow the \end{tabular} command.
\end{center}
\end{table}

\clearpage
%\appendix
\begin{table}
\begin{center}
\caption{Isotope ratios and fractional abundances of gaseous species at 10$^4$, 10$^5$ and 10$^6$ yr in the model with $n_{\rm H_2}=5\times10^3, 5\times10^4$ and $5\times10^5$ cm$^{-3}$}
\footnotesize
\begin{tabular}{cccccccc}
\tableline\tableline
        &            & 10$^4$ yr &                      & 10$^5$ yr &                      & 10$^6$ yr  &\\
Species & $n_{\rm{H_2}}$ (cm$^{-3}$) & $n_{{\rm i}}$/$n_{{\rm H}}$ & $^{12}$CX/$^{13}$CX & $n_{{\rm i}}$/$n_{{\rm H}}$ & $^{12}$CX/$^{13}$CX & $n_{{\rm i}}$/$n_{{\rm H}}$ & $^{12}$CX/$^{13}$CX\\
%        & &           &                      &           &                      &            &\\        
\tableline
CO & 5$\times 10^3$ & 6.0(-6) & 14& 2.8(-5)& 43& 2.5(-5)& 57\\
   & 5$\times 10^4$ & 1.7(-5) & 34& 1.6(-5)& 54& 1.3(-6)& 50\\
   & 5$\times 10^5$ & 1.2(-5) & 51&  2.0(-7)& 48& 1.2(-7)& 49\\
C & 5$\times 10^3$ &6.5(-5) &79 &2.6(-5) &80 &8.1(-9) &154 \\
   & 5$\times 10^4$ &4.0(-5) &79 &1.6(-8) &84 &3.0(-8) &77 \\
   & 5$\times 10^5$ &7.0(-7) &74 &9.7(-10) &65 &5.2(-10) &66 \\
HCO$^+$ & 5$\times 10^3$ & 3.4(-11) &88 & 2.3(-9) &51 &5.7(-9) &52 \\
   & 5$\times 10^4$ &4.9(-10)  &51 &1.9(-9) &45 &5.0(-10) &50 \\
   & 5$\times 10^5$ &2.7(-10) &42 &9.8(-11) &48 &7.3(-11) &49 \\
H$_2$CO & 5$\times 10^3$ &5.9(-10) &158 &1.7(-9) &74 &3.3(-9) &76 \\
   & 5$\times 10^4$ &2.9(-10) &76 &3.2(-10) &74 &2.4(-9) &68 \\
   & 5$\times 10^5$ &5.6(-11) &73 &1.6(-10) &56 &1.3(-10) &56 \\
CS & 5$\times 10^3$ &1.7(-8) &150 &2.6(-9) &83 &1.9(-10) &109 \\
   & 5$\times 10^4$ &1.4(-8) &109 &3.0(-10) &77 &5.1(-10) &75 \\
   & 5$\times 10^5$ &1.9(-9) &87 &4.4(-11) &72 &4.8(-11) &73 \\
CN & 5$\times 10^3$ &7.0(-9) &130 &8.0(-9) &83 &3.3(-9) &182 \\
   & 5$\times 10^4$ &4.6(-9) &83 &7.6(-10) &84 &7.7(-9) &77 \\
   & 5$\times 10^5$ &9.8(-10) &77 &6.9(-10) &67 &3.7(-10) &69 \\
CH & 5$\times 10^3$ &9.0(-9) &127 &8.0(-9) &78 &8.7(-10) &97 \\
   & 5$\times 10^4$ &5.0(-9) &78 &5.0(-10) &77 &1.5(-9) &75 \\
   & 5$\times 10^5$ &2.6(-10) &74 &1.1(-10) &71 &1.3(-10) &72 \\
CH$_2$ & 5$\times 10^3$ &6.2(-9) &102 &2.5(-9) &78 &8.2(-11) &88 \\
   & 5$\times 10^4$ &3.5(-9) &79 &1.2(-10) &75 &2.5(-10) &76 \\
   & 5$\times 10^5$ &6.7(-10) &73 &9.1(-11) &72 &1.1(-10) &73 \\
%C$_2$ & 5$\times 10^3$ &8.9(-9) &117 &5.8(-9) &83 &5.2(-10) &93 \\
%   & 5$\times 10^4$ &4.2(-9) &84 &8.0(-10) &76 &1.5(-9) &76 \\
%   & 5$\times 10^5$ &  & & & & & \\
HCN & 5$\times 10^3$ &3.6(-10) &151 &1.4(-8) &93 &5.3(-9) &180 \\
   & 5$\times 10^4$ &4.6(-8) &85 &1.5(-9) &96 &1.1(-8) &77 \\
   & 5$\times 10^5$ &4.1(-8) &77 &9.6(-10) &68 &4.1(-10) &69 \\
HNC & 5$\times 10^3$ &3.0(-10) &172 &1.3(-8) &96 &5.3(-9) &186 \\
   & 5$\times 10^4$ &3.6(-8) &88 &1.3(-9) &101 &1.1(-8) &77 \\
   & 5$\times 10^5$ &4.0(-8) &77 &9.2(-10) &67 &3.8(-10) &69 \\
\end{tabular}
\end{center}
\end{table}
\clearpage

\setcounter{table}{2}

\begin{table}
\begin{center}
\caption{continue}
\footnotesize
\begin{tabular}{cccccccc}
\tableline\tableline
        &            & 10$^4$ yr &                      & 10$^5$ yr &                      & 10$^6$ yr  &\\
Species & $n_{\rm{H_2}}$ (cm$^{-3}$) & $n_{{\rm i}}$/$n_{{\rm H}}$ & $^{12}$CX/$^{13}$CX & $n_{{\rm i}}$/$n_{{\rm H}}$ & $^{12}$CX/$^{13}$CX & $n_{{\rm i}}$/$n_{{\rm H}}$ & $^{12}$CX/$^{13}$CX\\
%        & &           &                      &           &                      &            &\\        
\tableline
HC$_3$N & 5$\times 10^3$ &4.8(-13) & 123&9.4(-10) &85 &1.2(-11) &99 \\
   & 5$\times 10^4$ &1.8(-9) &76 &1.4(-10) &73 &2.8(-11) &75 \\
   & 5$\times 10^5$ &9.5(-10) &71 &2.9(-13) & 73&9.4(-14) &71 \\
HC$_5$N & 5$\times 10^3$ &7.2(-14) &149 &7.2(-11) &83 &1.1(-12) &88 \\
   & 5$\times 10^4$ &4.3(-10) &76 &9.6(-12) &71 &2.2(-12) &74 \\
   & 5$\times 10^5$ &9.6(-11) &68 &1.6(-14) &73 &4.3(-15) &89 \\
CCH \footnotemark[1] & 5$\times 10^3$ &6.8(-9) & 136 &8.6(-9) & 80 &7.2(-10) &107 \\
   & 5$\times 10^4$ &3.3(-9) &78 &2.2(-9) &76 &8.3(-10) &76 \\
   & 5$\times 10^5$ &8.2(-10) &73 &2.7(-11) &71 &2.4(-11) &70 \\
CCS \footnotemark[1] & 5$\times 10^3$ &2.0(-10) &140 &1.0(-9) &81 &3.9(-10) &97 \\
   & 5$\times 10^4$ &1.3(-9) &83 &2.4(-10) &74 &2.3(-11) &75 \\
   & 5$\times 10^5$ &6.2(-10) &80 &2.9(-13) &81 &3.4(-13) &79 \\
C$_3$S & 5$\times 10^3$ &7.0(-12) &149 &1.0(-10) &82 &3.4(-12) &86 \\
   & 5$\times 10^4$ &1.1(-10) &82 &3.2(-11) &71 &2.5(-12) &74 \\
   & 5$\times 10^5$ &1.2(-10) &73 &1.8(-14) &79 &1.6(-14) &81 \\
\tableline
\end{tabular}
\tablecomments{(a) Isotope ratios for CCH and CCS are average isotope ratios of isotopomers.}
\end{center}
\end{table}

%% Any table notes must follow the \end{tabular} command.

\begin{table}
\begin{center}
\caption{Isotope ratios and fractional abundances of ice mantle species at 10$^4$, 10$^5$ and 10$^6$ yr in the model with $n_{\rm H_2}=5\times10^3, 5\times10^4$ and $5\times10^5$ cm$^{-3}$.}
\footnotesize
\begin{tabular}{cccccccc}
\tableline\tableline
        &            & 10$^4$ yr &                      & 10$^5$ yr &                      & 10$^6$ yr  &\\
Species & $n_{\rm{H_2}}$ (cm$^{-3}$) & $n_{{\rm i}}$/$n_{{\rm H}}$ & $^{12}$CX/$^{13}$CX & $n_{{\rm i}}$/n$_{{\rm H}}$ & $^{12}$CX/$^{13}$CX & $n_{{\rm i}}$/$n_{{\rm H}}$ & $^{12}$CX/$^{13}$CX\\
%        & &           &                      &           &                      &            &\\        
\tableline
CO ice & 5$\times 10^3$ &2.5(-7) &33 &2.8(-6) &37 &2.1(-5) &56 \\
   & 5$\times 10^4$ &2.7(-6) &35 &3.2(-5) &50 &2.6(-5) &51 \\
   & 5$\times 10^5$ &2.8(-5) &46 &4.0(-5) &48 &2.7(-5) &49 \\
CH$_3$OH ice & 5$\times 10^3$ &2.9(-8) &76 &7.2(-8) &80 &1.6(-6) &54 \\
   & 5$\times 10^4$ &2.0(-7) &81 &3.8(-7) &81 &2.2(-6) &54 \\
   & 5$\times 10^5$ &7.4(-7) &82 &6.2(-7) &81 &1.3(-6) &52 \\
HCOOH ice & 5$\times 10^3$ &1.1(-18) &129 &3.4(-12) &53 &2.5(-10) &59 \\
   & 5$\times 10^4$ &5.3(-13) &53 &2.9(-10) &46 &3.0(-10) &49 \\
   & 5$\times 10^5$ &4.8(-11) &45 &1.5(-10) &44 &1.1(-10) &44 \\
CO$_2$ ice & 5$\times 10^3$ &7.7(-10) &74 &6.1(-9) &81 &9.0(-8) &78 \\
   & 5$\times 10^4$ &9.3(-9) &81 &2.2(-7) &78 &1.7(-7) &76 \\
   & 5$\times 10^5$ &3.2(-7) &79 &5.2(-7) &78 &2.8(-7) &77 \\
H$_2$CO ice & 5$\times 10^3$ &6.5(-8) &75 &2.4(-7) &60 &1.4(-5) &54 \\
   & 5$\times 10^4$ &4.2(-7) &81 &9.1(-7) &80 &2.0(-5) &53 \\
   & 5$\times 10^5$ &1.3(-6) &80 &2.6(-6) &61 &1.5(-5) &50 \\
CH$_4$ ice & 5$\times 10^3$ &7.2(-7) &78 &8.2(-6) &82 &1.2(-5) &79 \\
   & 5$\times 10^4$ &8.1(-6) &81 &1.8(-5) &80 &2.0(-5) &78 \\
   & 5$\times 10^5$ &2.2(-5) &81 &2.2(-5) &81 &2.3(-5) &80 \\
\tableline
\end{tabular}
\end{center}
\end{table}

%% If the table is more than one page long, the width of the table can vary
%% from page to page when the default \tablewidth is used, as below.  The
%% individual table widths for each page will be written to the log file; a
%% maximum tablewidth for the table can be computed from these values.
%% The \tablewidth argument can then be reset and the file reprocessed, so
%% that the table is of uniform width throughout. Try getting the widths
%% from the log file and changing the \tablewidth parameter to see how
%% adjusting this value affects table formatting.

%% The \dataset{} macro has also been applied to a few of the objects to
%% show how many observations can be tagged in a table.

\clearpage

\end{document}